\documentstyle[preprint,aps]{revtex}
\begin{document}
\preprint{DAMTP/R-97/52}
\draft
\tighten
\title{BULK CHARGES IN ELEVEN DIMENSIONS}

\author{S. W. Hawking \footnote{E-mail:
    swh1@damtp.cam.ac.uk}
  and M. M. Taylor-Robinson \footnote{E-mail:
    mmt14@damtp.cam.ac.uk}}
\address{Department of Applied Mathematics and Theoretical Physics,
\\University of Cambridge, Silver St., Cambridge. CB3 9EW}
\date{\today}
\maketitle

\begin{abstract}
\noindent
{
Eleven dimensional supergravity has electric type currents arising 
from the Chern-Simon and 
anomaly terms in the action. However the bulk charge integrates to zero for 
asymptotically flat solutions with topological trivial spatial sections. We 
show that by relaxing the boundary conditions to generalisations of
the ALE and ALF boundary conditions in four dimensions one can obtain static
solutions with a bulk charge preserving between $1/16$ and $1/4$ of the
supersymmetries. One can 
introduce membranes with the same sign of charge into these backgrounds. This 
raises the possibility that these generalized membranes might decay quantum 
mechanically to leave just a bulk distribution of charge.
Alternatively and more probably, a bulk distribution of charge can
decay into a collection of singlely charged membranes. Dimensional 
reductions of these solutions lead to novel representations of extreme 
black holes in four dimensions with up to four charges. We discuss
how the eleven-dimensional Kaluza-Klein monopole wrapped around a
space with non-zero first Pontryagin class picks up an electric charge
proportional to the Pontryagin number. 
}
\end{abstract}
\narrowtext

\section{Introduction}
\noindent

The only bosonic fields in eleven dimensional supergravity are the 
metric and a three form potential $A$ for a four form field strength
$F$. The gauge 
symmetry is Abelian and the gravitino couples to the four form field strength 
rather than the potential. Thus it might seem that there were no
charged fields in the theory. However the action contains a 
Chern-Simons term \cite{CJS}
\begin{equation} 
S_{CS} \propto \int (A \wedge F \wedge F),
\label{chs}
\end{equation}
which implies that the divergence of the four form field strength is non zero
\begin{equation} 
d \ast F \propto F \wedge F. 
\label{div}
\end{equation}
In other words, there is an electric type bulk current 
$\ast (F \wedge F)$; the 
magnetic type bulk current is however zero since $dF=0$. This means that any 
magnetic charges $P_X=\int_X F$ where $X$ is a four cycle are purely 
topological and are the same for all homologous $X$. On the other hand the 
electric charges $Q_Y=\int_Y \ast F$ where $Y$ is a a closed seven surface
can have 
both topological and bulk contributions. There is also a contribution to the 
electric current and charge from the term in the bulk action that is needed to 
cancel anomalies from the world volume of the five brane \cite{DLM}.

One can integrate the electric current over an eight surface $K$ with 
$\partial K=Y$ to obtain the bulk contribution to the electric
charge. If $K$ were  
asymptotically Euclidean, one could conformally compactify by adding a
point at infinity. The Chern-Simons part of the electric current is 
conformally 
invariant. Thus its contribution to the charge at infinity must integrate to 
zero unless the fourth cohomology class $H^4(K,R)$ is non trivial. 
In particular, it will be zero for $K=R^8$ and 
the anomaly bulk contribution will also integrate to zero. This means that the 
ordinary asymptotically flat membrane cannot decay by losing its charge to a 
bulk distribution in space. 

On the other hand, a bulk electric charge can occur 
if $K$ is either not topologically trivial or not asymptotically Euclidean. As 
an example of the first kind, consider $S^4\times S^4$ and send a point to 
infinity by a conformal transformation to obtain an asymptotically Euclidean 
surface $ K$. One can choose the conformal factor, the scale of two extra
flat spatial dimensions and the four form to satisfy the time symmetric 
constraint equations. This would give time symmetric initial data for a 
solution with an electric charge that arose solely from bulk 
contributions. The classical evolution of such a solution would be 
collapse to a membrane.  However, we shall show 
that the situation 
may be different when $K$ is not asymptotically Euclidean but obeys some 
generalization of the ALE or ALF boundary conditions in four 
dimensions. In that case bulk charges can be classically stable though
one would expect that they may decay quantum mechanically into a 
collection of membranes which would have more 
phase space.

\bigskip

We will show that there are static solutions with bulk distributions of 
electric charge even when the topology of $K$ is $R^8$. Such solutions 
preserve a fraction of the supersymmetry provided that $K$ admits covariantly 
constant spinors. If the anomaly form confined to $K$ does not vanish, one 
must have such a bulk distribution of charge. Provided that $K$ admits a
self-dual harmonic $4$-form which is finite on the boundary $Y$ one can also
have contributions from a Chern-Simons bulk current. Solutions for 
which the Chern-Simons term is non-zero have not been much studied, but an 
elegant framework for incorporating such an $F$ in a fashion that preserves 
supersymmetry was developed in \cite{BB} and we extend this framework here. 

One can introduce generalized membranes with the same sign of charge into 
these backgrounds. This raises the 
possibility that membranes in these more general backgrounds could decay 
quantum mechanically into a bulk distribution of charge or vice
versa. Five branes
on the other hand can not decay in this way because there is no bulk 
magnetic current.

\bigskip

Several of the solutions which we construct have 
particularly interesting physical interpretations. 
For example, we find solutions describing eleven-dimensional 
Kaluza-Klein (KK) 
monopoles wrapped around topologically non-trivial worldvolumes; the anomaly
induces an electric type bulk charge. This result is as one would
expect: if one 
wraps any $p$-brane around a surface of non-zero first Pontryagin class 
one induces $(p-4)$-brane charges \cite{BSV}. 

When one wraps the KK monopoles around a torus or $K3 \times T^2$ one obtains 
extreme black holes in four dimensions, with up to four charges, some of
which arise from Chern-Simons and anomaly terms. The resulting 
four-dimensional spacetimes resemble those of multi-center 
black hole solutions, although the two types of solutions differ by 
the presence of pseudo-scalar fields in the former, which originate from the 
part of the $4$-form confined to the compact space. 
Black holes obtained by dimensional reduction of bulk 
distributions of charge have more singular horizons and 
higher temperatures than those with the same charge in membranes. 
This again suggests that bulk charges will decay quantum mechanically 
into membranes.

\section{Vacuum solutions}
\noindent

The bosonic part of the action of $d=11$ supergravity is given by 
\cite{CJS}
\begin{equation}
S_{11} = \frac{1}{2} \int d^{11}x \sqrt{g}  R - 
\frac{1}{2} \int (\frac{1}{2} F 
\wedge \ast F + \frac{1}{6} A \wedge F \wedge F)
\end{equation}
where $g_{MN}$ is the space time metric and $A$ is a 3-form with field 
strength  $F = dA$. We have set $\kappa^2$ to one.
The field strength obeys the Bianchi identity 
$dF = 0$ with its field equation being 
\begin{equation}
d\ast F = -\frac{1}{2} F \wedge F.
\end{equation}
In general, there will be gravitational Chern-Simons corrections to
this equation associated with the $\sigma$-model anomaly on the $d=6$ 
fivebrane \cite{DLM}. The corrected $5$-brane Bianchi identity takes the form 
\begin{equation}
d \ast F = - \frac{1}{2} F \wedge F + (2\pi)^4 \beta X_{8},
\end{equation}
where $\beta$ is related to the fivebrane tension by $T_{6} = \beta/(2\pi)^3$.
In all that follows we shall take $\beta= 1$ for simplicity. The $8$-form
anomaly polynomial can be expressed in terms of the curvature as
\cite{AW} 
\begin{equation}
X_{8} = \frac{1}{ (2\pi)^4} \lbrace - \frac{1}{768} ({\rm{Tr}} R^2)^2 + 
\frac{1}{192} ({\rm Tr} R^4) \rbrace, \label{af}
\end{equation}
and there is thence an additional term in the action of the form 
\begin{equation}
\Delta S_{11} = \frac{1}{2} \int A \wedge ( - \frac{1}{768} 
({\rm Tr} R^2)^2 + \frac{1}{192} ({\rm Tr} R^4)).
\end{equation}

\bigskip

We will start by looking for solutions of the form 
\begin{equation}
ds^2 = H(x^{m})^{-2/3} \lbrace ds^2(B^3) + ds^2 (B^8) \rbrace,
\end{equation}
where we take coordinates $x^{\mu}$, $\mu = 0,1,2$ on the Lorentzian 3-fold, 
and coordinates $x^{m}$, $m =3,.., 11$ on the Euclidean $8$-fold. 
We allow the scalar function to depend only on the latter and assume 
both that the anomaly form is 
non-trivial and that the Chern-Simons term does not vanish. Initially
we will consider vacuum solutions; that is, we do not include
membrane or fivebrane source terms. We will refer to the conformally
transformed metric as $\bar{g}_{MN}$, and associated covariant
derivative as $\bar{D}$.

It is natural to choose $B^{3}$ to be a symmetric space, and 
take $F$ to be of the form
\begin{equation}
F_{\mu\nu\rho n} = \pm \epsilon_{\mu\nu\rho} \partial_{n} f(x^{m}); 
\label{sgn}
\end{equation}
where $f(x^{m})$ is a scalar function that will be related to the
scale factor $H(x^m)$. We will also allow for a general $F$
on the $8$-fold, the form of which will be fixed by the field
equations. For clarity we express the $4$-form as the sum 
\begin{equation}
F = F_{1} + F_{2},
\end{equation}
where $F_{1}$ takes the form of (\ref{sgn}) and $F_{2}$ represents the
part of $F$ which is confined to the $8$-fold.

\bigskip 

The presence of the anomaly term in the Einstein equations makes it difficult 
to look for solutions of this form by solving the Einstein equations 
explicitly. Instead we look for a supersymmetric configuration 
satisfying this ansatz. 
Since the gravitino $\Psi_{M}$ vanishes in the background, the only 
non-trivial 
constraint on a supersymmetric solution is that for some Majorana
spinor $\eta$, variations of the gravitino vanish: 
\begin{equation}
\delta_{\eta} \Psi_{M} = D_{M} \eta - \frac{1}{288} (\Gamma_{M}^{PQRS} - 
8 \delta^{P}_{M} \Gamma^{QRS} ) F_{PQRS} \eta = 0,
\end{equation}
where $\Gamma_{M}$ is an eleven-dimensional gamma matrix. Our
conventions and notation for the gamma matrices are given in the Appendix. 

\bigskip

Solutions of this type, for which the $8$-fold is compact,  
were discussed in \cite{BB}; since the analysis of the field equations
changes little when the $8$-fold is non-compact
we give only a brief summary here. 
We decompose the eleven-dimensional gamma
matrices on the conformally transformed space by taking 
\begin{eqnarray}
\bar{\Gamma}_{\mu} &=& \gamma_{\mu} \otimes \gamma_{9}, \nonumber \\
\bar{\Gamma}_{m} &=& 1 \otimes \gamma_{m},
\end{eqnarray}
where $\gamma_{\mu}$ and $\gamma_{m}$ are gamma matrices of 
$B^3$ and $B^8$ respectively. $\gamma_{9}$ is the eight-dimensional
chirality operator which commutes with the $\gamma_{m}$ and satisfies
$\gamma_{9}^2 =1$. 

We then decompose the eleven-dimensional spinor $\eta$ as 
\begin{equation}
\eta = \epsilon \otimes \zeta(x^m),
\end{equation}
where $\epsilon$ is a three-dimensional anticommuting spinor, and 
$\zeta$ is a commuting eight-dimensional Majorana-Weyl spinor. Then the $\mu$
components of the gravitino equation reduce to the requirements that 
\begin{equation}
\bar{D}_{\mu} \epsilon = 0, \label{min}
\end{equation}
which implies that the symmetric three space $B^{3}$ must be
Minkowskian. In addition, $\zeta$ is of positive/negative chirality 
corresponding to the positive/negative signs in (\ref{sgn}), and the
function $f(x^m)$ is given by
\begin{equation} 
f(x^m) = H^{-1}(x^m).
\end{equation}
The resulting solutions take the form 
\begin{equation}
ds^{2} = H(x^m)^{-2/3}  ds^{2}(M^3) + H(x^m)^{1/3} ds^{2} (B^8). 
\end{equation}
One way to satisfy the $m$ components of the gravitino variation 
equation is to impose
the requirements that $B^{8}$ admits a complex structure and has
holonomy contained in $SU(4)$. The other is to assume that $B^{8}$ has
holonomy of precisely $Spin(7)$ which we will discuss below. In the
former case, the only non-zero
components of the $4$-form within the $8$-fold are the $F_{a\bar{b}
  c\bar{d}}$ components which must satisfy
\begin{equation}
F_{a\bar{b}c\bar{d}} J^{\bar{c} d} = 0, \label{kah}
\end{equation}
where $J$ is the K\"{a}hler form, and we use complex indices. 
On a compact K\"{a}hler space, we can express the solution for 
$F_{2}$ in terms of the harmonic $4$-forms $\omega^{i}_{4}$ as
\begin{equation}
F_{2} = \sum_{i=1}^{h_{11}} v^{i} \omega^{i}_{4}, \label{frs}
\end{equation}
where $h_{11}$ are components of the Hodge numbers. 
Quantisation of the magnetic charge imposes a constraint on the
$v^{i}$; defining a four-form $G$ which is related to $F_{2}$ by
normalisation factors, flux quantisation requires that \cite{W}
\begin{equation}
[G] - \frac{1}{4 (2\pi)^2} P_{1}(B^8) \in H_{4}(B^8,Z),
\end{equation}
where $P_{1}$ is the first Pontryagin class of $B^8$ and $[G]$ is the 
cohomology class of $G$. Note that we are defining the Pontryagin
classes without factors of $(2\pi)$. The existence of
spinors on $B^{8}$ implies that $P_{1}/(2\pi)^2$ is canonically divisible by
two. Depending on whether it is also canonically divisible by four,
$G$ will have integral or half-integral periods, and the coefficients 
$v^{i}$ will be integers or half-integers.  

\bigskip

Now the 32 component real spinor of the $d=11$ Lorentz group decomposes into 
representations of $SL(2,R) \times SO(8)$
\begin{equation}
{{\bf 32} \rightarrow (\bf{2,8_{s}}) \oplus (\bf{2,8_{c}}),}
\end{equation}
where ${\bf 8_{s}}$ and ${\bf 8_{c}}$ have opposite chiralities. 
If the holonomy is trivial, and $F$ is zero, then each of the 
${\bf 8}$s decomposes into eight singlets, and there are
$32$ covariantly constant spinors. If the holonomy is precisely 
$SU(4)$, then one of the spinor representations decomposes as
\begin{equation}
{ {\bf 8_{c}} \rightarrow \bf{6 \oplus 1 \oplus 1},}
\end{equation}
and there are thus a total of four covariantly constant spinors, 
of a defined chirality; only $1/8$ of the 
supersymmetry is preserved. If the holonomy breaks down further, a greater
fraction of the supersymmetry will be preserved. 

\bigskip

Given that this solution is obtained by requiring a fraction of the
supersymmetry to be preserved, it is natural to assume that the field
equations will also be satisfied. Let us first consider the equation
for the $4$-form.  Then the equation of motion for $F_{2}$ is 
\begin{equation}
d (\ast F_{2}) = - \frac{1}{2} F \wedge F = - F_{1} \wedge F_{2},
\end{equation}
where the conformal invariance of the field equation 
implies that there is no $X_{8}$
contribution. Now it was incorrectly stated in \cite{BB} that this
equation implies no further condition on the $4$-form 
than closure of $F_{2}$. 
Using the explicit form for $F_{1}$ (taking the sign in (\ref{sgn}) to
be positive) we find that 
\begin{equation}
d (H^{-1} \eta_{3} \wedge \ast_{8} F_{2}) = - \eta_{3} \wedge dH^{-1}
\wedge F_{2},
\end{equation}
where $\eta_{3}$ is the flat volume form on the transverse $3$-fold
and we take the dual $\ast_{8}$ in the Ricci-flat metric on the
$8$-fold. Since $F_{2}$ is harmonic, to satisfy this equation we must
take $F_{2}$ to be self-dual. If we reverse the sign in (\ref{sgn}), the
covariantly constant spinors on the $8$-fold must have the opposite
chirality, and the four-form $F_{2}$ must be anti-self-dual. Thus we
must include only self-dual/anti-self-dual forms in the summation 
(\ref{frs}). 

\bigskip

The field equation for $F_{1}$ does include an
anomaly term and imposes a constraint on the scalar function as
\begin{equation}
d( \ast_{8} d H) = - \frac{1}{2} F \wedge F +             
(2 \pi)^4 X_{8}, \label{sfe}
\end{equation}
where without loss of generality we have chosen the positive 
sign in (\ref{sgn}) and  $f(x^{m}) = H^{-1}(x^m)$. Note that we are 
again taking the dual in the Ricci-flat metric on the $8$-fold.  
When the $8$-fold is compact, there is a relationship between the
anomaly form defined in (\ref{af}) and the Euler class for any manifold
which admits a nowhere vanishing spinor \cite{IP}. For a nowhere vanishing 
positive chirality spinor to exist, the Euler class $e(B^8)$ must be
related to the first and second Pontryagin classes as 
\begin{equation}
8e(B^8) = 4P_{2} - P_{1}^2. \label{epn}
\end{equation}
One can show explicitly that this constraint is satisfied by all 
eight-dimensional Calabi-Yau manifolds.  Defining the Pontryagin
classes as 
\begin{equation}
P_{1} = - \frac{1}{2} {\rm Tr}R^2 \hspace{5mm} {\rm and} 
\hspace{5mm} P_{2} = - \frac{1}{4} {\rm Tr}R^4 
+ \frac{1}{8} ({\rm Tr}R^2)^2, \label{pont}
\end{equation}
it is apparent that that $X_{8}$ is related to the Euler class as 
\begin{equation}
X_{8} = - \frac{1}{4! (2\pi)^4} e(B^8) = -\frac{1}{4!} \rm{Pf}(R),
\end{equation}
where $\rm{Pf}(R)$ is the Pfaffian of the curvature form. 
The volume contribution to the Euler number is given by the integral
of $e(B^8)/(2\pi)^4$ over the manifold. Note that in \cite{BB} the
factors of $(2\pi)$  were omitted; one can easily verify that one
needs to include these factors when the Pontryagin classes are defined
without factors of $(2\pi)$. 

Integrating (\ref{sfe}) over a compact $8$-fold with no boundary, 
we find that 
\begin{equation}
\int_{B^{8}} F \wedge F = 2 (2\pi)^4 \int_{B^8} X_{8}.
\end{equation} 
Using the relationship to the Euler number we find the topological
constraint on the $8$-fold 
\begin{equation}
\int_{B^{8}} F \wedge F + \frac{(2\pi)^4}{12} \chi = 0, \label{con}
\end{equation}
where $\chi$ is the Euler number. Thus the constants $v^{i}$ are constrained, 
and the possible compactifications are restricted topologically. Since
the Euler number will not vanish for topologically non-trivial
compactifications, $F_{2}$ cannot vanish. One can obtain a natural
interpretation of this topological constraint in terms of the
quantisation law for the $4$-form \cite{W}. 

Whenever the anomaly form, and hence the Euler number, vanishes, the
$4$-form $F_{2}$ must be zero; we can then have nowhere vanishing spinors of
both chiralities on the $8$-fold, and in the vacuum solution there
will be up to $16$ conserved spinors of each chirality. 

If the anomaly does not vanish, one can also ensure that the net
electric charge vanishes by including membrane point singularities
which are localised within the $8$-fold. The constraints on the
$8$-folds and the membrane charges are discussed in \cite{SVW}. By
replacing the Chern-Simons term in (\ref{sfe}) by point membrane
contributions, one can obtain the metric for such solutions. 

\bigskip

We have discussed the analysis for positive chirality conserved spinors; let us
now consider an $8$-fold which admits covariantly constant spinors of
negative chirality. Then the ${\bf 8}_{s}$ spinor representation must admit a
decomposition containing singlets. 
From the above we know that we must include only
anti-self-dual forms in $F_{2}$ whilst from \cite{IP} we know that the
sign in (\ref{epn}) is reversed. Taking the minus sign in (\ref{sgn})
we find that 
\begin{eqnarray}
d(\ast_{8} dH_{-}) &=&  \frac{1}{2} F_{-} \wedge F_{-} - (2 \pi)^4
X_{(-)8} \\
&=&   \frac{1}{2} F_{-} \wedge F_{-} - \frac{1}{4! (2\pi)^4} e(B_{-}^8),
\end{eqnarray}
where we include minus signs to indicate that we are taking quantities
on an $8$-fold which admits negative chirality conserved spinors. Now
symmetry between the positive and negative chirality spinor solutions
implies that there exist solutions for which 
the warp factor $H(x^m)$ is the same for both. Then the Euler numbers
of the corresponding $8$-folds will be the same but 
both self-dual and anti-self-dual forms
and the ${\bf 8}_{s}$ and ${\bf 8}_{c}$ spinor representations will be
exchanged. In most of what follows we will assume that the conserved
spinors are of positive chirality. 

\bigskip

We said above that the other possible solution was an $8$-fold of
exceptional holonomy $Spin(7)$. Although we can certainly find such
solutions in the absence of an anomaly term \cite{PT}, 
one consequence of the constraints on $F_{2}$ 
is that compactification on a manifold of exceptional
holonomy $Spin(7)$ may not in general a solution of the field equations when we
include an anomaly term. If supersymmetry is to be preserved, the
harmonic $4$-form must satisfy the condition 
\begin{equation}
F_{mnpq} \tilde{\gamma}^{mnp} \zeta = 0, \label{sp7c}
\end{equation}
where $\tilde{\gamma}_{m}$ are the gamma matrices on the Ricci-flat
space $B^{8}$, as well as the topological condition (\ref{con}). 
As for the compact Calabi-Yau $8$-folds, a generic $Spin(7)$
manifold may not admit such a self-dual harmonic form satisfying 
the topological constraint. 

One can however show that the
unique $Spin(7)$ invariant self-dual $4$-form does satisfy this
condition; this is apparent if one defines it as \cite{GPP}
\begin{equation}
\omega_{mnpq} = \bar{\zeta} \tilde{\gamma}_{mnpq} \zeta,
\end{equation}
and uses the properties of Majorana-Weyl spinors and of
the gamma matrices. However to satisfy the
topological constraint one may need to include other harmonic forms in
$F_{2}$ which must also satisfy the condition above. 

\bigskip

Given this form of the solution, it is straightforward to confirm that the 
Einstein equations are also satisfied. Since the anomaly term is
related to a closed $8$-form, variations vanish within the $8$-fold,
but there will be contributions to the $11$-dimensional Einstein
equations whenever the anomaly form is non-trivial.  

\section{Non-compact vacuum solutions}
\noindent

Let us now suppose that the 8-fold is non-compact; the net 
electric charge need not vanish, since there is a boundary to the
$8$-fold. This implies that there are no topological constraints 
imposed by the anomaly term on the $4$-form.  

For general non-compact $8$-folds there may not be a
well-defined cycle over which we can integrate a $4$-form, and
hence magnetic charge is not well defined. This problem has been discussed
in the context of four dimensional Taub-Nut manifolds with dyonic
magnetic fields \cite{J}; one cannot define charge by integrating the 
relevant components of the Maxwell $2$-form over a sphere at infinity, since
the topology of surfaces of constant radius is non-trivial. One can 
only define charges by considering the motion of point charges in the
asymptotic region, or by taking analogies to asymptotically flat
space-times. Taking an $8$-fold which is the product of two such
Taub-Nut manifolds, the harmonic $4$-form cannot be integrated over a
$4$-cycle at infinity to give a quantisation condition. This will be a
generic problem in the manifolds of non-trivial holonomy that we
consider here. 

\bigskip

Although for the compact manifolds, the number of harmonic $4$-forms 
is given by the related Betti number $b_{4}$, for non-compact
manifolds we can include all harmonic $4$-forms whose norm is
finite, and which satisfy the constraints (\ref{kah}) or (\ref{sp7c}). 
This can include those not counted in the Betti number, since
they are non-zero on the boundary at infinity. In fact, harmonic
$4$-forms for which the magnetic behaviour is non-trivial cannot
vanish on the boundary at infinity. As an example, we can again 
consider Taub-Nut cross Taub-Nut; the fourth Betti number vanishes; 
yet there exists a harmonic form with finite norm which is non-zero 
on the boundary. 

\bigskip

In the compact case if the anomaly form is non-trivial one must
necessarily choose at least one of the $v^{i}$ to be non-zero for a 
solution to exist. In the non-compact case, the magnitude of the $4$-form is
arbitrary; one can choose the integers to be as large or small as one 
requires, and adjust the solution of the scalar equation accordingly. 
In particular, one can choose the $v^{i}$ to vanish, which has the
advantage of giving trivial magnetic behaviour. One can then certainly
have $Spin(7)$ solutions when the $8$-fold is non-compact, for which
one includes arbitrary amounts of the $Spin(7)$ invariant $4$-form. 

For positive chirality conserved spinors, we must still take the
four-form to be self-dual on the $8$-fold for a solution to
exist. For non-compact $8$-folds there is no topological obstruction 
to he existence of non-vanishing spinors, but the relationship between
the Pontryagin and Euler class (\ref{epn}) still holds, although the
integral of the Euler class of course gives only the volume
contribution to the Euler number. 

\bigskip

If $X_{8}$ is trivial, then we can choose $H(x^m)$ to be harmonic,
and, for vacuum solutions, to be a constant. The solutions we obtain this way 
are 
\begin{equation}
ds^2 = ds^2(M^3) + ds^2(B^8), 
\end{equation}
which are widely known. However, if the 
holonomy of $B^{8}$ is non-trivial, the anomaly term may not
vanish, and one cannot necessarily choose the scalar function to be 
constant. The physical interpretation is that there is a background charge
distribution over the $8$-fold. One would however expect that the anomaly 
term vanishes if the action of 
the holonomy group on one section of the tangent space of the $8$-fold
is trivial, and hence part of $B^{8}$ splits off as lines. Assuming a
fraction of the supersymmetry is preserved, the
anomaly will contribute only if the holonomy of $B^8$ is $Spin(7)$,
$SU(4)$, $Sp(2)$ or $Sp(1) \times Sp(1)$. Even if the holonomy is
contained in one of these groups, the anomaly can still vanish in
special cases, which we will discuss in \S 4. 

\bigskip

If the original $8$-fold is non-singular, the anomaly form $X_{8}$ 
should be smooth and continuous over the $8$-fold. Similarly, finiteness of the
energy will require that the Chern-Simons term $F \wedge F$ 
has no singularities. Thus the
background charge distribution implied by the equation for the scale
factor (\ref{sfe}) is smooth, non-zero, and finite at all points 
in the $8$-fold. This implies in turn that a non-singular smooth
solution for the scalar function will exist. 
We will give here three examples of such $8$-folds; the first has 
holonomy is $Sp(1) \otimes Sp(1)$, the second is Calabi-Yau whilst 
the third has holonomy of precisely $Spin(7)$. 

\bigskip

The simplest example of a non-trivial solution of this type is to 
take $B^{8}$ as Taub-Nut times Taub-Nut. 
One would expect more general $T^2$ invariant 
hyperK\"{a}hler metrics of this type, 
such as those constructed in \cite{GR}, to give solutions of a very
similar form. The
holonomy is $Sp(1) \otimes Sp(1)$, and one quarter of the supersymmetry is
preserved. There is a single harmonic $4$-form which is the wedge product
of the $2$-forms on the individual Taub-Nut manifolds. One can take the
metric to be  
\begin{equation}
ds^2 = H(x^m)^{-2/3} \lbrace ds^2(M^3) + H(x^m) 
(ds^2_{TN}(m_{1},x_{1}) + ds^2_{TN}(m_{2}, x_{2}) \rbrace 
\end{equation}
where we take the metric on each manifold to be 
\begin{equation}
ds^2_{TN}(m_1,x_{1}) = (1+ \frac{4m_{1}}{r_{1}})^{-1} 
(d\psi_{1} + \cos\theta_{1})^2 + (1+ \frac{4m_{1}}{r_{1}})
(dr_{1}^2 + r_{1}^2 d\Omega_{2}^2), 
\end{equation}
so that the $m_{i}$ are the nut parameters and $\psi_{i}$ is periodic
with period $16\pi m_{i}$. For simplicity we consider the
single-center metric on each manifold although this is an unnecessary
restriction. Since $B^{8}$ is a direct 
product of two $4$-folds, the (${\rm Tr} R^4$) term in the anomaly form
vanishes and we can show that
\begin{equation}
X_{8} \propto - \prod_{i} \frac{r_{i}}{(r_{i} + 4m_{i})^5} d\psi_{i} \wedge 
dr_{i} \wedge d\cos\theta_{i} \wedge d\phi_{i},
\end{equation}
where we will not need the constant of proportionality, but the
absolute sign is important. Since the volume contribution to the Euler
number of each Taub-Nut manifold is one \cite{EGH1}, $X_{8}$ integrates over
the $8$-fold to $-1/24$. 

If we include a non-trivial magnetic $4$-form, we must take it to be
\begin{eqnarray}
F &=& k^2 \prod_{i} \lbrace \frac{4m_{i}}{(r_{i} + 4m_{i})^2} 
d\psi_{i} \wedge dr_{i} + \frac{4m_{i}r}{(r_{i} +4m_{i})}
\sin\theta_{i} d\theta_{i} \wedge d\phi_{i} \nonumber \\
&& \hspace{15mm} - \frac{(4m_{i})^2}{(r_{i}+4m_{i})^2} \cos\theta_{i}
dr_{i} \wedge d\phi_{i} \rbrace, \label{tnf}
\end{eqnarray}
with $k$ a real constant, which has finite norm
\begin{equation}
\int_{B^{8}} F \wedge F = k^2 \prod_{i} (128 \pi^2 m_{i}^2).
\end{equation}
Note that the sign of $F \wedge F$ is positive, and $F$ is self-dual
on the $8$-fold. The magnetic ``charge'' can be expressed as 
\begin{equation}
\int_{\prod_{i} S^2_{r_{i} \rightarrow \infty}} F = k \prod_{i} (16
\pi m_{i}).
\end{equation}
Now the equation for the scalar function $H(x^m)$ (\ref{sfe}) can be
expressed in coordinate notation (in terms of the
Ricci-flat metric on Taub-Nut cross Taub-Nut, $\tilde{g}_{mn}$) as 
\begin{equation}
\tilde{D}_{n} \partial^{n} H(x^m) = g(r_{i}), \label{pn}
\end{equation}
where $g(r_{i})$ is a negative definite function on $B^{8}$. 
Since the
function depends only on the radii, the charge distribution is
delocalised in the toroidal and angular directions. It is important to
note that however positive or negative we choose the magnetic charge
to be the function in (\ref{pn}) will always be negative. 
If one ignores the anomaly and $4$-form terms, then the field
equations are solved provided that $H(x^m)$ is harmonic, as was 
found in \cite{GGPT}. The anomaly term will give only a small
correction 
to the scalar function at infinity, but cannot be neglected. 

Supposing one defines the charge as 
\begin{equation}
q = \int_{B^8} d \ast F,
\end{equation}
then the contribution to $q$ from the background of (\ref{pn}) is
negative since the charge density is negative definite throughout the
manifold. The scalar function $H(x^m)$ satisfies a Poisson equation,
with the negative charge density being concentrated about the origin
of the two manifolds, and decaying at infinity. The form of the
operator on the product manifold makes it difficult to solve the
equation explicitly, but we would expect that there exists a
non-singular solution of the form 
\begin{equation}
H(x^m) = 1 + h(r_{i}),
\end{equation}
where $h(r_{i})$ is a function which is positive definite throughout
the manifold $B^{8}$, peaks at a finite value at the origin $r_{i} = 0$ and
asymptotically vanishes. 

\bigskip

We have been referring to this as a vacuum solution, but 
it is better interpreted in terms of eleven-dimensional 
Kaluza-Klein monopoles. When $m_{1} = 0$ the solution can be
described in terms of KK $6$-branes \cite{Ro} located at the origin
$r=0$. The anomaly and Chern-Simons terms vanish and 
one half of the spacetime supersymmetry is preserved by the solution.
If the $6$-branes are located in the 
$(123456)$ plane, then the condition for unbroken supersymmetry is
\cite{BKOP}  
\begin{equation}
\Gamma_{0123456} \eta = \eta, \label{onc}
\end{equation}
which again gives us $16$ conserved spinors. If we then add
$6$-branes located in the $(12789(10))$ plane only spinors satisfying 
\begin{equation}
\Gamma_{012789(10)} \eta = \eta, \label{ons}
\end{equation}
preserve supersymmetry. One half of the spinors satisfying
(\ref{onc}) will also satisfy this condition, and so one quarter of
the supersymmetry is preserved, as we found above. 

Suppose one dimensionally reduces the two monopole solution along
closed orbits of the Killing vector $\partial_{\psi_{2}}$. Then the
$6$-branes lying in the $(123456)$ plane are reduced to the $D6$-branes 
of type IIA theory \cite{T1}, whilst those lying in the $(12789(10))$ plane 
are reduced to IIA KK monopoles. Since the $D6$-brane charge is
quantised, the ``nut'' charges will determine the number of
$D6$-branes in ten dimensions. 

The effective ten-dimensional solution then describes $D6$-branes 
intersecting with IIA monopoles, a
configuration preserving $1/4$ of the supersymmetry. The presence of a
non-zero anomaly form implies that one should have electric
charge corrections to such a solution. As mentioned in the
introduction, whenever a $p$-brane is wrapped around a surface of
non-zero first Pontryagin class, one expects that the $p$-brane picks
up a $(p-4)$-brane charge. Our explicit construction of the spacetime
solution demonstrates the eleven-dimensional origin of the 
membrane charge of the $D6$-brane induced when it wraps a topologically 
non-trivial space. 

\bigskip

If one chooses $B^8$ as Taub-Nut cross any $4$-fold of $Sp(1)$ 
holonomy $B^4$, then the anomaly form is given by 
\begin{equation}
X_{8} = \frac{1}{4 (2\pi)^4 4!} [P_{1}(B^4) \wedge P_{1}(TN)].
\end{equation}
Integrating over the $8$-fold we find that the induced electric 
charge is
\begin{equation}
q = (2 \pi)^4 \int_{B^8} X_{8} = \frac{(2 \pi)^2}{48} p_{1}(B^4), 
\label{gc}
\end{equation}
where $p_{1}(B^4)$ is the first Pontryagin number of the $4$-fold.
Now in \cite{SVW} the value of the charge induced on a $p$-brane 
wrapped around a surface of non-zero first Pontryagin class
was given as $1/48$ of the first Pontryagin number. Allowing for 
normalisation differences in the $4$-form and the Pontryagin classes,
our result is consistent. 

\bigskip

Our second example is the generalised Eguchi-Hanson solution in 
eight dimensions. 
This manifold was first discussed in \cite{CA}, \cite{GZ} 
and the metric was given in the form that we 
shall use here in \cite{MMT}. It is an ALE Ricci-flat K\"{a}hler
manifold, which hence has holonomy $SU(4)$. The form of the metric is 
\begin{equation}
ds^2_{8} 
= \frac{dR^2}{(1-\frac{a^8}{R^8})} + \frac{R^2}{16}((1-\frac{a^8}{R^8})
(d\tau + A)^2 + R^2 ds^2(CP^3),
\end{equation}
where the metric $g_{ij}$ on $CP^3$ is chosen with the scale 
$R_{ij} = 8 g_{ij}$
and $dA$ is the K\"{a}hler form on the complex projective
space. The behaviour of this solution is analogous to that of the 
more familiar four dimensional solution; in particular, 
the radial coordinate runs
between $a$ and infinity, and the $CP^3$ fixed point set of the isometry
$\partial_{\tau}$ allows us to calculate the Euler number as four
using the Lefschetz fixed point theorem.

The anomaly form for this manifold is 
\begin{equation}
X_{8} = -  \frac{14}{\pi^4} 
\frac{a^{32}}{R^{33}} dR \wedge d\tau \wedge \eta_{6},
\end{equation}
where $\eta_{6}$ is the volume form on the complex projective
space. By calculating the surface contribution to the Euler number,
one can then verify that the anomaly form gives the correct volume
contribution. 

Assuming that the harmonic form vanishes the warp factor equation is then 
\begin{equation}
\tilde{D}_{n} \partial^{n} H(x^m) = -896 \frac{a^{32}}{R^{40}}, \label{hf}
\end{equation}
which has the regular solution 
\begin{equation}
H(x^m) = 1 + 28 \lbrace \frac{1}{6R^6} + \frac{a^8}{14 R^{14}}
+ \frac{a^{16}}{22 R^{22}} + \frac{a^{24}}{30 R^{30}} \rbrace. \label{slf}
\end{equation}
One can also take the harmonic form $F_{2}$ to be non-vanishing; if $F_{2}$ is
proportional to the (self-dual) canonical four form then there will be an
additional contribution to (\ref{hf}) which is also negative definite
and proportional to $1/R^{16}$. Solving for the scalar function
then gives an additional $1/R^6$ term in (\ref{slf}). 

\bigskip

As a third example, we mention an $8$-fold with $Spin(7)$ holonomy 
which was discussed in \cite{GPP}. The metric takes the form of a 
quaternionic line bundle over a 4-sphere
\begin{equation}
ds^2_{8} = \alpha^2(r) dr^2 + \beta^2(r) (\sigma^{i} - A^{i})^2 + \gamma^2(r)
ds^2_{4},
\end{equation}
where $ds^2_{4}$ is a suitably scaled metric on the base $4$-sphere and
$\sigma^{i}$ are left-invariant one-forms on the $SU(2)$ fibres of the
bundle over $S^{4}$. The functions $\alpha$, $\beta$ and $\gamma$ are
given by
\begin{equation}
\alpha^2 = (1-r^{-10/3})^{-1}; \hspace{2mm} 
\beta^2 = \frac{9}{100}r^2 (1-r^{-10/3})^{-1}; 
\hspace{2mm} \gamma^2 = \frac{9}{20} r^2.  
\end{equation}
Asymptotically this solution
tends to the metric on the cone
\begin{equation}
ds^2 = d\rho^2 + \rho^2 ds_{7}^{2}(S^7),
\end{equation}
where the metric on the seven-sphere is a homogeneous ``squashed'' Einstein
metric. Given the curvature tensor for such
a solution calculated in \cite{PP}, we can calculate the anomaly form;
again it describes a smooth negative charge distribution. Inclusion of
an $F_{2}$ term satisfying the constraint (\ref{sp7c}), such as
the $Spin(7)$ invariant $4$-form, simply modifies the
warp factor and increases the positive charge background. 

\section{Generalised membranes}
\noindent

Given vacuum solutions asymptotic to $M^3 \times B^{8}$ which preserve
some or all of the supersymmetry, it is natural to ask whether we can
include membranes. The backgrounds discussed above remain
supersymmetric solutions if we
include contributions to $H(x^m)$ which are harmonic on the
$8$-fold. Point singularities are naturally interpreted as the
positions of parallel membranes, and these membranes do not 
necessarily break any more of the supersymmetries. 

\bigskip

It is worth considering here the nature of harmonic solutions
on the $8$-fold. For the standard membrane on $R^8$, the natural
choice of harmonic function describes a single membrane localised at 
what can be chosen to be the origin of the $8$-fold. The equation
satisfied by the scalar function is 
\begin{equation}
\partial^{n} \partial_{n} H(x^m) = - \alpha \delta^{8}(x^m),
\end{equation}
where the delta function integrates over the manifold to give one. 
Then we choose 
\begin{equation}
H(x^m) = 1 + \frac{1}{6 V_{7}} \frac{\alpha}{r^6},
\end{equation}
where $V_{7}$ is the volume of the seven sphere. This choice of scalar
function gives the familiar membrane solution of \cite{DS}. 
The charge can be defined as 
\begin{equation}
q = \int_{R^8} d \ast F =  - \alpha,
\end{equation}
as expected. Evidently inclusion 
of further point singularities in the harmonic function simply
describes additional parallel membranes. Such solutions preserve $1/2$ of the
supersymmetry. Note that the scalar function is positive definite when
$\alpha$ is positive. 

Now let us consider harmonic functions on the product of two Taub-Nut
manifolds. Suppose we look for a solution which depends neither on the
position in one of the Taub-Nut manifolds nor on the circle direction
in the other. An appropriate solution is given by 
\begin{equation}
\delta H(x^m) \propto \frac{\alpha}{r_{1}}.
\end{equation}
where our notation refers to the change in the scalar function induced
by the inclusion of a point singularity. 
This describes a membrane of negative charge 
which is localised at the origin of one of
the Taub-Nut manifolds, but which is delocalised in the other
manifold, and along the circle direction. Such a solution preserves
only $1/4$ of the supersymmetry, provided of course that we include an
appropriate anomaly term. 

The ten-dimensional interpretation is as a delocalised membrane
contained within a $D6$-brane and IIA monopole. 
As we would expect the charge determined by
this harmonic function diverges, although the charge per unit volume
of the second Taub-Nut manifold is finite; that is, the divergence is
caused by delocalising the membrane over a non-compact manifold. 

If we take the function to be the sum of two harmonic
functions on the individual $4$-folds, the resulting solution will
represent two parallel membranes, each of which is localised at the
origin of one manifold, but delocalised in the other manifold. Such a 
solution still preserves $1/4$ of the spacetime supersymmetry. 

\bigskip

If we want the membrane to be localised at the origin of each
Taub-Nut manifold, then we need to look for a solution to
\begin{equation}
\delta \tilde{D}_{n} \partial^{n} H(x^m) = - \alpha \delta^{8}(x^m),
\label{egs}
\end{equation}
where the delta function implies that the membrane is localised
both in the circle directions, and at the radial origin of each
$4$-fold. Note that we have chosen the sign so that the change in the
scalar function is positive definite throughout the manifold. There
are several reasons for this choice: $\alpha$ must be positive if 
$H(x^m)$ is not to pass through zero and if the mass of the membrane 
is to be
positive. The solution to the equation above does not have a simple analytic
form, and we will not discuss the explicit 
solution. We would expect the scalar function to be mildly singular at the
membrane location, although the singularity may behave similarly
to the horizon of the ordinary membrane \cite{GHT}. 

\bigskip

We can find an explicit solution for the other $8$-folds 
discussed in the previous section. For example, for the generalised
Eguchi-Hanson solution, the change to the scalar function obtained by
solving (\ref{egs}) is 
\begin{equation}
\delta H(x^m) = \frac{3\alpha}{32 \pi^4 a^{6}} \lbrace 
\ln(\frac{R^2+a^2}{R^2-a^2}) + 2 \tan^{-1} (\frac{R^2}{a^2}) - \pi
\rbrace.
\end{equation}
Then the scalar function falls off as $1/R^{6}$ at infinity, and
diverges logarithmically at the origin $R=a$. Unlike the ordinary
membrane, for which the spacetime approaches the regular manifold
$AdS_{4} \times S^7$ \cite{GHT} at the membrane, the size of the complex
projective spaces will blow up logarithmically as we approach the
membrane.  Since the source is distributed over a six-dimensional
complex projective space, one might regard such branes as being
in some sense eight-dimensional.   

\bigskip

Let us now consider whether we can interpret these types of solution 
as membranes. Since the equation for the warp factor 
on $B^{8}$ takes the form (\ref{pn}), a solution including  
point singularities in $H(x^m)$ can be interpreted as localised 
membranes within a background charge distribution. 
Now we have found that such point singularities must have a definite 
charge, implied by taking $\alpha$ to be positive in (\ref{egs}). 
If such a solution is to 
represent a membrane solution, the choice of signs in the spacetime
equations must be consistent with the choices made to satisfy the membrane 
field equations. 

\bigskip

The bosonic sector of the membrane action \cite{DS} is given by
\begin{eqnarray}
S_{M} = T \int d^3\xi( -\frac{1}{2} \sqrt{-\gamma} \gamma^{ij}
\partial_{i} X^{M} \partial_{j}X^{N} g_{MN} + \frac{1}{2}
\sqrt{-\gamma} \nonumber \\ 
\pm \frac{1}{3!} \epsilon^{ijk} \partial_{i}X^{M} \partial_{j}X^{N} 
\partial_{k}X^{P} A_{MNP} ), \label{wz}
\end{eqnarray}
where $T$ is the membrane tension, and $\gamma_{ij}$ is the metric on 
the membrane world-volume, and $\xi^{i}$ are world-volume coordinates.
$X^{M} = (X^{\mu}, Y^{m})$ are the spacetime coordinates, with $\mu =
0,1,2$ and $m=3,..,11$.  
There is a correction to the energy momentum tensor arising from the membrane
source term 
\begin{equation}
\delta T_{MN} =  T \int d^3\xi \sqrt{-\gamma} \gamma_{ij} 
\partial^{i}X_{M} \partial^{j}X_{N} \frac{\delta^{11}(x-X)}{\sqrt{-g}}.
\end{equation}
As in \S 2 we have taken $\kappa^2 =1$. The corresponding equation of motion 
for the four-form then gives a correction to the scalar function
\begin{equation}
\delta \tilde{D}_{n}\partial^{n} H(x^m) = - T 
\int d^3\xi \epsilon^{ijk} \partial_{i}X^{0}\partial_{j}X^{1} 
\partial_{k} X^{2} \delta^{11}(x - X).
\end{equation}
From the membrane action, we have the membrane field equations 
\begin{eqnarray}
\partial_{i} (\sqrt{-\gamma} \gamma^{ij} \partial_{j}X^{N} g_{MN}) 
&+& \frac{1}{2} \sqrt{-\gamma} \gamma^{ij} \partial_{i} X^{N} \partial_{j} 
X^{P} g_{NP} \nonumber \\
 &\pm & \frac{1}{3!} \epsilon^{ijk} \partial_{i}X^{N} \partial_{j}X^{P} 
\partial_{k} X^{Q} F_{MNPQ} = 0, \\
\gamma_{ij} &=& \partial_{i} X^{M} \partial_{j} X^{N}
g_{MN}. \nonumber 
\end{eqnarray}
Now if one takes the static gauge choice and solution
\begin{eqnarray}
X^{\mu} &=& \xi^{\mu}; \nonumber \\
Y^{m} &=& {\rm constant},
\end{eqnarray}
one can verify that with the choice of $F_{MNPQ}$ corresponding to
positive chirality conserved spinors the
membrane field equations are satisfied provided that we choose the
negative sign of the Wess-Zumino term in (\ref{wz}), and vice versa.  
The correction to the scalar function satisfies 
\begin{equation}
\delta \tilde{D}_{n}\partial^{n} H(x^m) = - T \delta^{8}(x^{m}).
\end{equation}
Comparison with the source term (\ref{egs}) then implies that 
$\alpha = T$, which is analogous to the relationship for the
ordinary membrane \cite{DS}.  

\bigskip

We then need to determine the number of supersymmetries that are
preserved when we include membrane point singularities. Preservation
of the world-volume supersymmetries requires that the spinor $\eta$
must satisfy the condition
\begin{equation}
\Gamma \eta = \eta,
\end{equation}
where we have taken the sign to be negative in the Wess-Zumino 
term (\ref{wz}), and 
\begin{equation}
\Gamma \equiv \frac{1}{3!\sqrt{-\gamma}}  \epsilon^{ijk}
  \partial_{i}X^{M} \partial_{j} X^{N} \partial_{k}X^{P} \Gamma_{MNP},
\end{equation}
For our solutions we have $\Gamma = 1 \otimes \gamma_{9}$, so that the
preserved world-volume supersymmetries are of the same chirality as the
preserved spacetime supersymmetries. Thus the membrane does not break
any more spacetime supersymmetries than the vacuum solution, and our
solutions can indeed be interpreted as a membranes with $T^2$ invariant
hyperK\"{a}hler, Calabi-Yau and $Spin(7)$ transverse spaces preserving
$1/4$, $1/8$ and $1/16$ of the supersymmetries respectively. 

For the monopole solutions, one can see why the addition
of a membrane in the $(12)$ plane to a $(123456)$ and $(12789(10))$
brane configuration does not break any more supersymmetries. The above
condition on the spinor does not impose any more restrictions that the
two conditions $(\ref{onc})$ and $(\ref{ons})$. 

\bigskip

There are several points to make about these generalised membrane
solutions. All the solutions which we have discussed describe 
positive/negative charge membranes within a positively/negatively
charged background. At first one might think that this indicates 
a possible instability of such
membranes, in which the membranes decay into a background with a
greater Chern-Simons term. The ordinary membrane could not of course 
decay in such a way, since there exists no harmonic $4$-form on $R^8$ which is
finite at infinity. However it seems more likely that the bulk charges
would decay quantum mechanically into membranes, a point that will be
illustrated in the following section. 

\bigskip

We should also mention the differences between the generalised
membranes discussed here and those considered in
\cite{DLPS}. The motivation for the construction of the latter was the
observation that the solution for the ordinary membrane is asymptotic to 
$AdS_{4} \times S^{7}$ at the membrane location. Since there exist 
other Einstein $7$-folds which admit Killing spinors, one might expect that 
there exist analogous membranes which are asymptotic to $AdS_{4}
\times B^{7}$ where $B^{7}$ is a more general positive curvature 
Einstein manifold. The form of these solutions is 
\begin{equation}
ds^2 = H(r)^{-2/3} ds^2(M^3) + H(r)^{1/3} [dr^2 + r^2 ds_{7}^2(B^7)],
\end{equation}
where $H(r)$ is harmonic, and membranes contribute $1/r^6$ terms to
the scale factor. Evidently as for the ordinary membrane such
solutions are non-singular at the membrane location. The Ricci-flat
metric on the $8$-fold has holonomy contained in $Spin(7)$ when one 
chooses the manifold $B^7$ suitably, such as the squashed seven-sphere
solution of \cite{ADP}.  Thus these membranes constitute 
one class of the solutions considered here, although we have allowed the
metric to take a more general form. 

Note that such a background is not complete in the absence of
membranes; unless the metric on $B^7$ is the round metric on the
sphere, there will be conical singularities at the origin $r=0$. In
fact, the $Spin(7)$ $8$-folds of this type, Ricci-flat metrics on
cones, first constructed by \cite{Bry} were all incomplete. The
squashed seven-sphere solution mentioned above is closely related to
the $Spin(7)$ manifold discussed in \S 3; in the latter we smooth out
the cone, with the singular ``vertex'' at $r=0$ being replaced by a
smoothly-embedded bolt.  

Neither anomaly nor Chern-Simons terms were included in the
analysis of \cite{DLPS}. Since the $8$-fold is not flat, one might
expect there to be corrections to the scalar function from an anomaly
term; however, the form of the metric on the $8$-fold indicates that
the volume contribution to the Euler number vanishes, and no
corrections are needed. In addition one cannot find a self-dual
$4$-form on the $8$-fold which integrates to give a finite charge. 

\bigskip

All of the $8$-folds which we have considered admit at least a circle
isometry group. As in \cite{GGPT}, we could 
dimensionally reduce the vacuum and membrane solutions to ten
dimensions, and then apply duality transformations to obtain new
solutions. Supersymmetry is not preserved by the dimensional
reduction unless the Killing spinors are also invariant under the
action of the isometry, a condition which is non-trivial for general 
$Spin(7)$ $8$-folds. However supersymmetry will certainly be preserved
if we take the $8$-fold to be hyperK\"{a}hler and $T^2$ invariant as in
\cite{GGPT}. We will not consider such dimensional reductions here,
although in the next section we will consider dimensionally
reduced solutions of a different type.

\section{Modified Kaluza-Klein monopole solutions}
\noindent

So far we have mostly been interested in solutions which are
manifestly eleven-dimensional for which the anomaly form is non
vanishing. In this section we will consider the effects of including
anomaly and Chern-Simons forms for solutions which can best be
interpreted in lower dimensions.  

When we compactify the solutions we need to be careful about the
quantisation condition on the $4$-form. The vacuum solutions we
consider are of the same form as those in \S 3 except that 
the $8$-fold is conformal to the product $B_{1} \times B_{2}$ where
$B_{1}$ is compact (and one or more directions in $B_{2}$ is wrapped
around a circle). 

Preservation of any of the spacetime supersymmetry requires that 
$B_{1}$ is a torus or $K3$. The quantisation condition on the 
four-form is evidently trivial for the former, and,
since the first Pontryagin class of $K3$ is canonically 
divisible by four, $G$ must have integral periods \cite{CS} 
on $K3$ also. One cannot find a self-dual four-form on $B^8$ of
non-zero period over $K3$ which is finite on the boundary, and so $G$
has vanishing period over the compact $4$-fold in the solutions
considered here. 

\bigskip

Directly dimensionally reducing an eleven-dimensional KK $6$-brane gives a  
$D6$-brane in ten dimensions and wrapping this 
$D6$-brane around a torus gives an
extreme four-dimensional black hole carrying a $U(1)_{M}$ magnetic
charge, which preserves one half of the supersymmetry.
The formal temperature of the black hole as defined by the surface
gravity is infinite, and thus it has
a naked singularity which is protected by an infinite mass gap \cite{BL}. 

Now for a single KK $6$-brane with a flat transverse space, say in the
$(123456)$ directions, the anomaly form necessarily vanishes. If the transverse
space $(3456)$ is non-compact, then one cannot find a Chern-Simons
form $F_{2}$ on the $8$-fold which would give a finite charge. 
If however we wrap the KK $6$-brane around a four torus, we can find 
such a form. Since in this case
$F_{1}$ must be non-zero, only spinors of positive chirality on the
$8$-fold preserve the supersymmetry, and hence $1/4$ of the
supersymmetry is preserved. The eleven-dimensional interpretation of
such a solution is as a generalised monopole solution with electric
charge corrections.

If we further compactify the $(12)$ directions on a two-torus, 
and take the circle direction in the Taub-Nut to be small, we again 
obtain an extreme black hole in four dimensions. That is, the 
eleven-dimensional solution is
\begin{eqnarray}
ds^2 &=& H(r)^{-2/3} [-dt^2 + ds^2(T^2)] + H(r)^{1/3} ds^2(T^4) 
 + H(r)^{1/3} \lbrace h(r) ds^2(R^3) + \nonumber \\
&& + h(r)^{-1} (d\psi + 4m \cos\theta d\phi)^2  \rbrace, \label{upt}
\end{eqnarray}
where 
\begin{equation}
h(r) = (1 + \frac{4m}{r}); \hspace{10mm} H(r) = (1 + 
\frac{4mk^2}{(r+4m)}). \label{ypg}
\end{equation}
The four-form is given by 
\begin{eqnarray}
F_{1} &=& dt \wedge \eta(T^2) \wedge dH(r)^{-1}; \label{tyf} \\
F_{2} &=& \omega_{T^4} \wedge \omega_{TN}, \nonumber 
\end{eqnarray} 
where $\omega_{T^4}$ is the (constant) self-dual $2$-form on $T^4$ and
$\omega_{TN}$ is the self-dual $2$-form on Taub-Nut, given in
(\ref{tnf}). Then the charge in eleven dimensions is given by
\begin{equation}
\int_{B^8} d (\ast F) = 64 \pi^2 m V_{4} (4m k^2),
\end{equation}
where $V_{4}$ is the volume of the $T^4$. Using the standard
ansatz for dimensional reduction of the eleven-dimensional solution 
\cite{CY}, we
find that the effective four-dimensional solution (in the Einstein
frame) is
\begin{eqnarray}
ds^2_{E} &=& - h(r)^{-1/2} H(r)^{-1/2} dt^2 + h(r)^{1/2} H(r)^{1/2}[dr^2
+ r^2 d\Omega_{2}^{2}]; \nonumber  \\ 
F^{E} &=& dt \wedge dH(r)^{-1} = \frac{4mk^2}{(r+4m(1+k^2))^2} dt
\wedge dr \label{4d} \\
F^{M} &=& 4 m \sin \theta d\theta \wedge d\phi; \nonumber
\end{eqnarray}
with the other effective fields in four dimensions being the dilaton,
and a pseudo-scalar originating from $F_{2}$. The notation for the
gauge fields indicates that the charges come from different gauge
groups, $U(1)_{M}$ and $U(1)_{E}$. 

\bigskip

From the form of the metric, the effective four-dimensional
solution seems to describe an extreme $U(1)_{M}$ black hole of mass 
$m$ with (singular) horizon at $r=0$, plus an extreme 
$U(1)_{E}$ black hole of mass $m k^2$ with (singular) horizon at $r= -4m$. 
We can however rewrite the solution as 
\begin{equation}
ds^2_{E} = - (1+ (1+k^2)\frac{4m}{r})^{-1/2} dt^2 +
(1+ (1+k^2)\frac{4m}{r})^{1/2} [dr^2 + r^2 d\Omega_{2}^{2}],
\end{equation}
which looks like the metric for an extreme black hole carrying only
{\it one} charge, with horizon at $r=0$ and mass $m(1+k^2)$. 
Since the black hole again has a formal temperature
which is infinite, it is protected by an infinite mass gap even though
only one quarter of the supersymmetry is preserved and there are two
non-zero charges. 

\bigskip

We could also add a membrane to the KK monopole solution; the form of
the solution is the same as in (\ref{upt}) except that the scalar
function $H(r)$ is now defined as 
\begin{equation}
H = 1 + \frac{q}{r},
\end{equation}
where we choose $F_{2}$ to vanish and delocalise the charge over the
internal torus as required by the Kaluza-Klein ansatz. The charge in 
eleven dimensions is given by
\begin{equation}
\int_{B^8} d (\ast F) = 64 \pi^2 m V_{4} q, 
\end{equation}
whilst the effective fields in four dimensions are as in (\ref{4d})
with the only other scalar field being the dilaton. This solution
describes a $U(1)_{M} \times U(1)_{E}$ black hole with horizon at 
$r=0$ and singularity at $r = -q$ or $r= -4m $ depending on the
relative magnitudes of $q$ and $m$. Since the formal temperature of
the black hole is finite, it is protected from excitations 
by a finite mass gap. Again one quarter of vacuum supersymmetry is
preserved. 

For given electric charge, the masses of both 
types of black hole solutions are of course the same, although 
the temperatures 
and singularity structure differ. Actually the four-dimensional
solutions differ only in the background fields, and the location of
the electric black hole. Although the most interesting membrane plus
$D6$-brane intersections are those for which the branes are localised 
at the same point in the transverse space, one can find 
a multi-center solution in which
the membrane in ten dimensions is located at $r < 0$ whilst the
$D6$-brane is located at $r=0$. The resulting four-dimensional black
hole solution differs from (\ref{upt}) only by the absence of the 
pseudo-scalar field.

\bigskip

The most general eleven-dimensional solution of this type will have
non-zero $k$ and $q$, preserving one quarter of the supersymmetry. The
four-dimensional solution can be interpreted in terms of 
electric black holes with singular horizons located at $r=0$ and $r=-4m$ and
magnetic black holes with singular horizons located at $r=0$. The effective
solution has a non-singular horizon at $r=0$ and a singularity 
at some $r < 0$. 

\bigskip

It is also interesting to consider KK $6$-branes wrapped
around $K3 \times T^2$. In the supergravity theory one can interpret
such a solution in four dimensions as an extreme 
magnetic black hole preserving one quarter of the supersymmetry. Since
the anomaly form on $K3 \times$ Taub-Nut is non-zero, one needs to
include a warp factor in the solution. So the metric is 
\begin{eqnarray}
ds^2 &=& H(r)^{-2/3} [ -dt^2 + ds^2(T^2)] + H(r)^{1/3} ds^{2}(K3) +
H(r)^{1/3} \lbrace h(r) ds^2(R^3) \\
&& + h(r)^{-1} (d\psi + 4m \cos \theta d\phi)^2 \rbrace,  \nonumber 
\end{eqnarray}
where $h(r)$ is given in (\ref{upt}). Calculation of the anomaly form
then implies that the scale factor satisfies
\begin{equation}
\tilde{D}_{n} \partial^{n} H(x^m) = - 12 c \frac{(4m)^2}{(r+4m)^6},
\end{equation}
where $c$ is a real constant which could be calculated. This equation
is straightforward because the Euler class of $K3$ is proportional to
the volume form. Then the scalar
function is  
\begin{equation}
H(r) = 1 + \frac{c}{4m} \lbrace \frac{1}{(r+4m)} + \frac{4m}{(r+4m)^2}
+  \frac{(4m)^2}{(r+4m)^3} \rbrace,
\end{equation}
and the associated electric charge is given by
\begin{equation}
\int_{B^{8}} d \ast F = 16 \pi^2 c V_{K3} = (2 \pi)^4,
\end{equation}
where $V_{K3}$ is the volume of the $K3$ and the Euler number of $K3$
($\chi = 24$) is used in calculating the latter equality.  
Since the first Pontryagin number of $K3$ is $48 (2\pi)^2$ one can 
also obtain this result from (\ref{gc}).

\bigskip

Reduction to four dimensions again gives us a black hole 
carrying two charges, which preserves one quarter of the
supersymmetry, but is singular with a formally infinite temperature. The
presence of the anomaly form causes the black hole to carry an
electric charge as well as a magnetic charge. 

One can include several different Chern-Simons forms $F_{2}$ in these
solutions; choosing
$F_{2}$ as in (\ref{tyf}), the $2$-form on $K3$ must be self-dual. There
are three distinct self-dual $2$-forms on $K3$, but the only way in
which the $2$-form, $\omega_{2}$, will contribute to the 
equation for motion is via the wedge product 
$\omega_{2} \wedge \omega_{2}$. As the latter is cohomologous to
the unique harmonic volume four-form of $K3$, we know that 
\begin{equation}
\omega_{2} \wedge \omega_{2} = C \eta_{K3} + d \omega_{3},
\end{equation}
where $\eta_{K3}$ is the volume form and $C$ is a constant. 
The requirement that $\omega_{2}$ is self-dual implies
the exact term vanishes \cite{CS}, and  
so we can find solutions for which the correction to the scalar
function is of the form (\ref{ypg}). Chern-Simons
terms modify the electric charge although
the temperature of the black hole remains infinite. Inclusion of
membranes wrapped around the two-torus gives a four-dimensional black
hole with a finite temperature in the extremal state. 

\bigskip

More generally, of course, we could wrap further membranes about
holomorphic cycles in the torus or $K3$. 
One can have a $(2 \perp 2 \perp 2) \parallel 6_{KK}$ configuration in
which each of the membranes is wrapped around a torus. What is novel
about our solutions is that one can also include
Chern-Simons contributions to the $4$-form from the self-dual $4$-form
on the $8$-fold transverse to each membrane. With suitable choices of
charge signs, the configuration still preserves $1/16$ of the vacuum
supersymmetry. 

The effective four-dimensional black hole as usual has
four charges and a finite horizon area, but the metric will be non-standard
and pseudo-scalar fields will be non-zero. Depending on the
relative sizes of the charges, there may be another horizon inside the 
horizon at $r=0$ before one reaches a physical singularity. 

Evidently the usual further generalisations of intersecting brane
solutions are possible. 
Instead of taking the membranes to intersect over a point, one can
choose them to be at general angles, with the Chern-Simons forms
chosen appropriately. 
One can also wrap additional membranes about holomorphic cycles in $K3$; 
solutions of this kind were discussed in \cite{CH}. 
The effective four-dimensional single center solutions then
preserve one eighth of the supersymmetry and describe
extreme black hole with three $U(1)$ charges. 

A general feature of all these solutions is that 
black holes obtained by dimensional reduction of bulk 
distributions of charge have more singular horizons and 
higher temperatures than those with the same charge in membranes. 
This suggests that these bulk charges will decay quantum mechanically 
into membranes.

\bigskip

We should briefly mention ``non-extreme'' generalisations of these
solutions. For the torus solution, one can make the 
eleven-dimensional monopole solution
non-extreme by taking the metric to be of the form 
\begin{eqnarray}
ds^2 = - f(r) dt^2 + ds^2(T^6) + h(r)[ f(r)^{-1} dr^2 + r^2
d\Omega^2_2] \nonumber \\
+ h(r)^{-1} (d\psi + \sqrt{4m(4m+\mu)} \cos\theta d\phi)^2, \label{nex}
\end{eqnarray}
where $h(r)$ is defined in (\ref{upt}) and 
\begin{equation}
f(r) = (1 - \frac{\mu}{r}).
\end{equation}
In the eleven-dimensional solution there is a regular null horizon 
at $r=\mu$, but the surface $r=0$ is singular. As we take the limit 
of $\mu \rightarrow 0$, the temperature of the monopole diverges.  
The four-dimensional interpretation of this
solution is a magnetic black hole with outer horizon at $r=\mu$ and inner
(singular) horizon at $r=0$. The charge is proportional to 
$\sqrt{4m(4m+\mu)}$ whilst the mass is $m + \mu/2$, and the
temperature diverges as we take the extremal parameter to zero.  

One can generalise the single-center membrane solutions in the same
way, following the ansatz of \cite{CT}; these non-extreme configurations
should be regarded as ``bound-states'' rather than as intersections 
of non-extreme branes. However a four-dimensional extreme solution 
which is a multi-center black hole system does not have a static
non-extreme generalisation; the fields become time dependent, and the
black holes approach the same location. It would be interesting to
determine what happens to KK monopoles carrying Chern-Simons
charges when one adds a little energy to the BPS solutions.

\section{Five-brane solutions}
\noindent

If one considers a five-brane within an eleven-dimensional background
for which the anomaly form is non-zero, then one has to take account of
electric charge corrections. For example, 
corrections will be required for the generalised
five-brane solutions discussed in \cite{GKT}. The simplest solution to 
consider is that for a single five-brane 
\begin{eqnarray}
ds_{11}^2 &=& {\cal{F}}^{-1/3}(ds^2(M^2) + ds_{4}^{2}(B_{1})) + 
{\cal{F}}^{2/3} (ds_{4}^2(B_{2}) + dz^2); \nonumber \\
F &=& \ast_{2} d{\cal{F}} \wedge dz, 
\end{eqnarray}
where we take the dual in the last equation on the manifold $B_{2}$. 
$M^2$ is $2$-dimensional Minkowski space, whilst the manifolds $B_{i}$
must be Ricci-flat to satisfy the field equations. 
$\cal{F}$ is a harmonic function on this manifold, and the magnetic 
charge is given by integrating $F$ over the boundary of $B_{2}$ cross 
the line. Again point singularities in $\cal{F}$ represent localised
five-branes. 

\bigskip

Preservation of any of the spacetime supersymmetry 
requires that the holonomy of
each of the $B_{i}$ is contained in $SU(2) \cong Sp(1)$. Usually one
assumes that the manifolds are flat, and one half of the background 
supersymmetry is then preserved in the $5$-brane solution.  

If $B_{1}$ has trivial holonomy, but $B_{2}$ has holonomy $Sp(1)$, 
then the vacuum solution preserves $1/2$ of the supersymmetry, and the
$5$-brane solution preserves $1/4$ of the supersymmetry, with suitable
choice of charge sign. If both
of the manifolds have holonomy $Sp(1)$ the vacuum solution preserves
$1/4$ of the supersymmetry, but the $5$-brane solution preserves only
$1/8$ of the supersymmetry. This follows from the fact that the vacuum
solution preserves eight Killing spinors, four of each chirality  
on the six-dimensional manifold $M^2 \times B_{2}$. If $\cal{F}$ has
point singularities, then only spinors of one particular chirality on the
world-volume preserve the supersymmetry. 

In both of the $B_{i}$ are non-trivial, however, the
anomaly form does not vanish. Using the conformal invariance of the field
equation for the $4$-form, we know that the anomaly polynomial is
transverse to the conformally flat space. Note that if only one of
the manifolds has non-trivial holonomy conformal invariance implies
that the anomaly form vanishes. 

Since a $D4$-brane wrapped around a space for
which the first Pontryagin class does not vanish picks up an induced
$0$-brane charge and the $M5$-brane reduces to the $D4$-brane on 
double dimensional reduction, one might wonder from where this charge
originates in eleven dimensions. In fact,
this charge originates from the self-dual $3$-form field propagating
on the worldvolume of the $M5$-brane. However we do not need to use the 
worldvolume fields of the five-brane in what follows and can ignore
the non-zero value of this field. 

\bigskip

To find a solution which takes account of the anomaly, it is
natural to add a correction of the type discussed in the previous
sections. The corrected five-brane solution will take the form
\begin{eqnarray}
ds_{11}^2 &=& {\cal{F}}^{-1/3} H^{-2/3} \lbrace (ds^2(M^2) + H
ds_{4}^{2}(B_{1})) \nonumber \\
&& + {\cal F}^{2/3} (H ds_{4}^2(B_{2}) + dz^2) \rbrace; \nonumber \\
F &=& \ast_{2}d{\cal F} \wedge dz + \eta_{M^2 \times R} \wedge 
dH^{-1}, 
\end{eqnarray}
where $\eta$ is the volume form on the flat space. ${\cal F}$ remains a
function which is harmonic on $B_{2}$, whilst $H$ satisfies the 
equation (\ref{sfe}), with additional 
point singularities representing membranes. 

One can verify that the addition of such scalar function terms to the
metric and to the $4$-form does not break any additional
supersymmetries, provided that one chooses charge signs
appropriately. Even if one includes point singularities in $H(x^m)$
representing membranes, $1/8$ of the supersymmetry is preserved. 

The physical interpretation of this result is that a $5$-brane wrapped
around a space of non-trivial holonomy, with a non-flat transverse space
receives electric charge corrections when one takes account of the anomaly. 
Note that just as for the single membrane one add an arbitrary amount
of self-dual harmonic $4$-form on the $8$-fold; this will not affect 
the amount of supersymmetry that is preserved. 

\bigskip

Furthermore the solution representing two
$5$-branes intersecting on a string, which was discussed in \cite{GGPT}, 
must in general be corrected to 
\begin{eqnarray}
ds_{11}^2 &=& ({\cal F}_{1}{\cal F}_{2})^{2/3} H^{-2/3} 
\lbrace ({\cal F}_{1}{\cal F}_{2})^{-1}
ds^2 (M^2) + {\cal F}_{1}^{-1} H ds_{4}^{2}(B_{2}); \nonumber \\
&& \hspace{15mm} + {\cal F}_{2}^{-1} H ds_{4}^{2}(B_{1}) +dz^2
\rbrace \label{gec} \\
F &=& (\ast_{1}d{\cal F}_{1} + \ast_{2} d{\cal F}_{2}) 
\wedge dz \pm \epsilon_{M^2  \times R} \wedge dH^{-1}; \nonumber 
\end{eqnarray}
with $H(x^m)$ satisfying (\ref{sfe}). Here
$M^{2}$ is Minkowski space, and $B_{i}$ are Ricci-flat four-dimensional
manifolds with holonomy contained in $Sp(1)$. 
In the last line, $\ast_{i}$ implies that 
the duals are taken on the manifolds $B_{i}$. 
Such a solution is the same as in \cite{T}, except that the definition
of $H(x^m)$ includes both anomaly and Chern-Simons terms. 

In the absence of an anomaly term, the field equations are 
satisfied provided that the functions $H_{i}$ are harmonic on the
manifolds $B_{i}$. If the $B_{i}$ are flat, then each fivebrane
preserves $1/2$ of the supersymmetry, and the overlap preserves $1/4$
of the supersymmetry. Inclusion of point singularities in $H(x^m)$
gives a solution preserving $1/8$ of the supersymmetry. 

If only one of the $B_{i}$ is flat, then we obtain a solution when 
$H(x^m)$ is harmonic. If $H(x^m)$ is constant, then $1/8$ of the
supersymmetry is preserved. Inclusion of point singularities in
$H(x^m)$, i.e. membranes wrapped around the conformally flat space,
does not affect the amount of supersymmetry which is preserved. 
If both of the $B_{i}$ are hyperK\"{a}hler, the scalar function is no
longer harmonic, but $1/8$ of the supersymmetry is preserved, whether
or not we include membranes, provided that we choose the signs of
charges suitably. Again one could add an arbitrary amount of 
$4$-form on the $8$-fold, without breaking any more supersymmetry. 

\bigskip

Although the $B_{i}$ can be any manifolds of $Sp(1)$
holonomy, one obtains the most interesting physical interpretations 
for $T^4$, $K3$ and Taub-Nut manifolds. 
Suppose that both of the $B_{i}$ are Taub-Nut manifolds. Then our 
general solution (\ref{gec}) describes two five-branes intersecting a membrane
over a string. Each five-brane is parallel to one KK $6$-brane and
intersects the other over the common string. One could wrap further
membranes around holomorphic cycles in the Taub-Nut manifolds to
obtain solutions preserving smaller fractions of the supersymmetry. 

If one of the $B_{i}$ is a torus, and the other is
Taub-Nut, then compactification to four dimensions of the single five-brane
plus monopole solution gives us a black hole with two magnetic charges
which preserves $1/4$ of the supersymmetry. Addition of a Chern-Simons
term leads to a black hole carrying three charges
which preserves $1/8$ of the supersymmetry, but which still has a
finite temperature. If we include membranes, the four-dimensional 
interpretation of the solution is as a black hole carrying three
charges, preserving $1/8$ of the supersymmetry, which has zero
temperature. Obviously wrapping the $5$-branes around $K3 \times T^2$ gives
analogous black hole solutions. 

\bigskip

Solutions describing three overlapping five-branes are known; for
example one can have the fivebranes 
all overlapping on a string, and each pair overlapping on a
three-brane \cite{GKT}. 
As one would expect, the solutions generically preserve $1/8$ of the
background supersymmetry. However the anomaly form necessarily
vanishes, since the three-brane spaces must be conformally Ricci 
flat, and thence Riemann flat. Even if one wraps the branes around
tori, there exists no finite Chern-Simons form. 
More generally, the anomaly form can only be non-zero when we have 
two transverse four dimensional manifolds with non-trivial holonomy,
or an eight-dimensional manifold with non-trivial holonomy. 

There is a more general class of five-brane solutions related to the
vacua $B^3 \times B^8$ where the holonomy of a manifold conformal to 
$B^{8}$ is a larger subgroup of $Spin(7)$. One can wrap a five-brane around
$M^2 \times Y_{4}$, where $Y_{4}$ is a $4$-fold within the
$8$-fold. A related discussion wrapping branes about cycles within
such manifolds can be found in \cite{CH} and \cite{Bo}. 
With suitable choice of the $4$-fold one may
preserve a fraction of the supersymmetry. Since the anomaly form is
non-vanishing, one gets electric charge corrections, and the
resulting five-brane solutions are the generalisation of 
that given above. 

There is also a class of solutions obtained by including
Brinkmann waves. One can include a wave to any intersection involving
at least a common string. Thus one could for example 
add a wave to a KK $6$-brane 
parallel to a five-brane intersecting a membrane along a
string. Wrapping around a torus or $K3 \times T^3$, the boosting 
gives us a four charge black hole in four dimensions. Chern-Simons and
anomaly terms give corrections to the mass, charge and
singularity structure of the resulting black hole.  

\appendix
\section*{Conventions} \label{app}
\noindent

The different type of indices that we use are as follows. $M = 0,..,10$
represent eleven-dimensional space-time indices. $\mu = 0,1,2$
represent three-dimensional space-time indices. $m = 3,..,10$
represent eight-dimensional space-time indices. We use $A = 0,..,10$
to indicate eleven-dimensional tangent space indices.  
When referring to $8$-folds admitting a complex structure, we use
indices $a, \bar{a}$ where $a = 1,..,4$. 

The $d=11$ Dirac matrices $\Gamma_{M}$ satisfy 
\begin{equation}
\lbrace \Gamma_{M}, \Gamma_{N} \rbrace = 2 g_{MN}, 
\end{equation}
where $g_{MN}$ has signature $(-,+,..,+)$. $\Gamma_{M_{1}..M_{n}}$ is
the anti-symmetrised product
\begin{equation}
\Gamma_{M_{1}..M_{n}} = \Gamma_{[M_{1}}...\Gamma_{M_{n}]},
\end{equation}
where the square bracket implies a sum over $n!$ terms with a $1/n!$
prefactor. The chirality operator is defined by 
\begin{equation}
\gamma_{9} = \frac{1}{8!} \epsilon_{mnpqrstu} \gamma^{mnpqrstu},
\end{equation}
whilst our definition of the Hodge star is
\begin{equation}
\ast(dx^{m_{1}} \wedge... \wedge dx^{m_{p}}) = \frac{1}{(d-p)!}
  \epsilon^{m_{1}...m_{p}} {}_{m_{p+1}...m_{d}} dx^{m_{p+1}..m_{d}}.
\end{equation}  
For further conventions and identities applying to the Dirac matrices,
see the Appendix of \cite{BB}. A further convention used in the
derivation of (\ref{min}) is that $\epsilon_{012} = 1$ on $M^3$.

\end{document}